\documentclass[10pt,twocolumn,preprintnumbers,superscriptaddress,nofootinbib,aps,prd,floatfix]{revtex4-1}
\pdfoutput=1
\usepackage{enumerate}
\usepackage{gensymb}
\usepackage{amsmath,amssymb}
\usepackage{mathrsfs}
\usepackage{graphicx}
\usepackage{slashed}
\usepackage{xspace,slashed}
\usepackage{hyperref}
\hypersetup{colorlinks=true, citecolor=blue, urlcolor=blue, linkcolor=blue}
\usepackage[normalem]{ulem}
\usepackage{subfigure,orcidlink}
\usepackage{array}
\usepackage[autostyle]{csquotes}
\usepackage{multirow,array}
\usepackage{float}
\usepackage{hyperref}
\hypersetup{colorlinks=true, citecolor=blue, urlcolor=blue, linkcolor=blue}
\usepackage{tabularray}
\UseTblrLibrary{booktabs}
\usepackage[absolute,overlay]{textpos}

\begin{document}

\title{Inelastic Dark Matter at LZ from Radiative Dirac Neutrino Mass Paradigm}

\author{Pankaj Borah${\orcidlink{0000-0003-2715-271X}}$}
\email{pankajborah@iitg.ac.in}
\affiliation{Department of Physics, Indian Institute of Technology Guwahati, Assam 781039, India}

\author{Satyabrata Mahapatra\,\orcidlink{0009-0009-9592-4786}}
\email{satyabrata@iitgoa.ac.in}
\affiliation{School of Physical Sciences, Indian Institute of Technology Goa, Ponda-403401, Goa, India.}

\author{Newton Nath~\orcidlink{0000-0002-0592-0020}}\email{nnath.phy@iitbhu.ac.in}
\affiliation{Department of Physics, Indian Institute of Technology (BHU), Varanasi 221005, India}

	
\date{\today}
\begin{abstract}
The LUX-ZEPLIN (LZ) experiment has reported a high-energy nuclear recoil candidate,
LZ230616, at $E_R = 248 \pm 23_{\rm stat} \pm 23_{\rm sys}\,$keV, which is
difficult to reconcile with elastic scattering of halo dark matter (DM). Endothermic
inelastic scattering offers a natural explanation, but it rests on a nucleon
coupling that is off-diagonal rather than diagonal.
We show that this feature arises automatically within a
radiative Dirac neutrino mass framework. The Standard Model is extended by
three right-handed neutrinos (RHNs), a pair of vector-like neutral fermions for each generations, and a scalar sector comprising
an inert doublet and a real singlet, governed by a
$\mathcal{Z}_2 \times \mathcal{Z}_4$ symmetry.
The $\mathcal{Z}_2$ symmetry stabilizes the DM, while the $\mathcal{Z}_4$ symmetry forbids the tree-level Dirac Yukawa coupling together with all renormalizable Majorana mass terms, thereby allowing the generation of Dirac neutrino masses at the one-loop level through a softly broken scalar trilinear coupling `$\kappa$'.
Since the neutral dark-sector fields can be expressed in terms of real scalar mass eigenstates, the $Z$ boson couples to them purely off-diagonally and elastic 
$Z$-mediated scattering is absent identically rather than simply suppressed. The
singlet--doublet mixing induced by `$\kappa$' governs both the inelastic rate and
the one-loop Dirac neutrino mass, which therefore vanish as
$\kappa \to 0$. We find that DM masses in the few hundred GeV to TeV range, with
splittings of $\mathcal{O}(340-360)\,$ keV, simultaneously reproduce the
observed relic abundance, account for the LZ event and
yield neutrino masses of the correct order, while respecting elastic
direct-detection limits and all relevant theoretical and experimental constraints. A substantial part of the surviving parameter space lies within reach of DARWIN.
\end{abstract}

\maketitle	
\section{Introduction}

The present Universe is composed of approximately one-quarter dark matter (DM) by energy density~\cite{Planck:2018vyg}, while its underlying particle nature remains one of the major challenges to address in astroparticle physics. Since DM is non-baryonic, the highly celebrated Standard Model (SM) of particle physics is unable to account for its existence, motivating the search for new particles and interactions beyond the SM. Among several strategies deployed to probe its identity, direct detection experiments of DM have made spectacular progress in recent times. Dual-phase xenon time projection chambers
led by LUX-ZEPLIN (LZ)~\cite{LZ:2025igz}, XENONnT~\cite{XENON:2023cxc} and PandaX-4T~\cite{PandaX-4T:2021bab}, now probe
spin-independent DM--nucleon cross sections at the level of
$10^{-48}\,\mathrm{cm}^2$ for weakly interacting massive
particle (WIMP) masses of a few tens of
GeV and are beginning to approach the irreducible neutrino
background.
Recently, the LZ experiment has reported an event in an extended nuclear-recoil search, corresponding to a recoil energy of $ E_R = 248 \pm 23_{\rm stat} \pm 23_{\rm sys}\ {\rm keV}, $
in a $2.84$ tonne-year exposure \cite{LZ:2026axp}. The event lies in a region where the expected background is small, and its high recoil energy makes it particularly interesting for DM scenarios involving non-standard recoil spectra, including inelastic DM scattering. The background-only hypothesis is disfavored at a global significance of $2.6\sigma$, with a maximum local significance of $3.4\sigma$ across the models considered. The kinematic characteristics of this event hint at dynamics beyond the standard elastic scattering paradigm. 
Several recent studies have explored possible explanations for the observed LZ event~\cite{Freese:2026sga,Lou:2026idn,Yin:2026jnn,DiMauro:2026ldr,Visinelli:2026kgt,Yamashita:2026ump,Rodd:2026tyn,Du:2026guj,McCabe:2026crm,Jeesun:2026vzo,Unwin:2026rdp,Baer:2026fpy,Das:2026uyy,Alhazmi:2026efz,Ahmed:2026qjg,Du:2026lpa,Kannike:2026qyl,Bose:2026ndd,Bandyopadhyay:2026gjw,Borah:2026zwf,Okada:2026eol,Bisal:2026khf,Cheung:2026byg,Elahi:2026vlm,Zhu:2026dag,Lee:2026jxl,Aghaie:2026vsu,Khan:2026nwp,Kotlarski:2026pep,Langhoff:2026ujr,He:2026hqz,Kumar:2026lgi,Nguyen:2026lui,Heikinheimo:2026kwp,Chattaraj:2026fxn, Dent:2026bji,Bose:2026szs}.

Inelastic or endothermic DM~\cite{TuckerSmith:2001hy} provides a natural explanation of such a signal. If the dark sector contains two
nearly degenerate states $X_1$ and $X_2$ with a small splitting
$\delta \equiv m_{X_2} - m_{X_1}$, and if the coupling to the
nucleon is {off-diagonal}, then the  up-scattering $X_1 N \to X_2 N$, is
kinematically allowed only above
$
  v_{\rm min} \;=\; \frac{1}{\sqrt{2 m_N E_R}}
  \left( \frac{m_N E_R}{\mu_{X N}} + \delta \right) ,
$
with $\mu_{X N}$ the reduced mass.
For $\delta \sim \mathcal{O}(100\,\mathrm{keV})$, only particles in the
high-velocity tail of the halo distribution carry enough kinetic energy to
up-scatter. The rate is correspondingly suppressed, and the surviving recoils are
confined to a narrow band at high $E_R$, a region in which elastic scattering
predicts essentially no signal. This explanation, however, rests on two ingredients that are ordinarily imposed by
hand. The first is a pair of states with
mass-splitting lying in the narrow keV window which is fixed by the halo kinematics. The
second is a nucleon coupling that is off-diagonal rather than diagonal, since any
unsuppressed diagonal coupling would generate elastic scattering at a level
already excluded.
In the framework developed below, the second of these is not an assumption but an exact consequence of the symmetry that renders the neutrino a Dirac particle, while the strength of that off-diagonal coupling is fixed by the same parameter that generates the neutrino mass.
The splitting itself remains a free parameter of the scalar potential, with which the recoil energy of the event is reproduced.

Besides DM, neutrino masses provide another compelling piece of evidence for physics beyond the SM (BSM), with their discovery recognized by the 2015 Nobel Prize in Physics~\cite{Super-Kamiokande:1998kpq, SNO:2001kpb}. In addition to their tiny masses, the nature of neutrinos, whether they are Dirac or Majorana particles, remains an open problem, continuing to motivate extensive theoretical and experimental studies. The simultaneous existence of DM and neutrino masses, along with their nature, therefore provides a strong motivation to explore BSM frameworks that can address these phenomena within a common theoretical framework. There have been several models that have made successful attempts to explain the origin of tiny neutrino masses and DM within a unified framework. Among them, the ``scotogenic" mechanism~\cite{Ma:2006km} is a particularly well-known minimal extension of the SM, in which Majorana neutrino masses are generated radiatively, and the stability of DM is ensured by an imposed discrete $\mathcal{Z}_2$ symmetry. 

Neutrinoless double beta decay experiments that could probe the Majorana nature of neutrinos have so far provided no conclusive evidence for their Majorana nature~\cite{Schechter:1981bd}. This leads to an interesting possibility that neutrinos are Dirac particles and motivates further theoretical and experimental investigations of their Dirac nature.  Recently, numerous studies have explored such Dirac scotogenic scenarios~\cite{Ma:2016mwh,Bonilla:2018ynb,Dasgupta:2019rmf,Guo:2020qin,CentellesChulia:2024iom,Kumar:2025cte,Paul:2026snc,Kang:2026lgr}. These frameworks generate neutrino mass at the one-loop level while ensuring that neutrinos remain Dirac particles along with the stable DM candidate. 

\begin{figure}[t!]
\centering
\includegraphics[width=\linewidth]{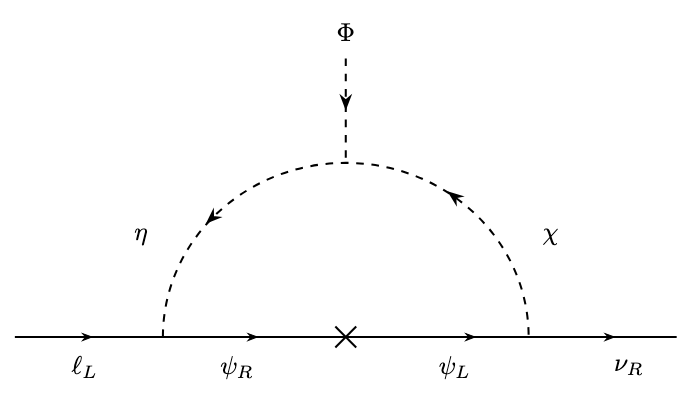}
\caption{One-loop Dirac neutrino masses.}
\label{fig:Nu-Dirac1loop}
\end{figure}

In this work, we develop a formalism that accounts for the tiny neutrino masses while preserving their Dirac nature and, at the same time, provides an inelastic DM candidate that could potentially explain the recent LZ event. To achieve our goals, we utilize $\mathcal{Z}_2\times \mathcal{Z}_4$ symmetry, in addition to the SM symmetries that lead to Dirac neutrino mass, as shown in Fig.~\ref{fig:Nu-Dirac1loop}.
Symmetries are incorporated in such a way that the Dirac neutrino mass remains forbidden at the tree-level and at dimension-4, while it can be generated at the one-loop level \footnote{Some recent studies that address one-loop realization of dimension-4 operator along with the Dirac nature of neutrinos can be found in~\cite{Ma:2016mwh,Bonilla:2018ynb,Ma:2019yfo, CentellesChulia:2024iom,Okada:2026pek}.}.
Importantly, we introduce a soft symmetry breaking term that violates the $\mathcal{Z}_4$ symmetry and induces mixing among the dark scalars. This mixing plays a crucial role in generating Dirac neutrino masses at the one-loop level. On the other hand, the discrete $\mathcal{Z}_2$ symmetry ensures the stability of the DM candidate, thereby allowing it to account for the observed relic abundance of the Universe.

The dark sector of our construction consists of an inert scalar doublet together
with a real singlet, and this combination is what makes the LZ event accessible.
An inert scalar doublet on its own already offers the possibility of inelastic
DM as the quartic coupling $(\Phi^\dagger\eta)^2$ splits its neutral component into
a CP-even and a CP-odd real state, and since real scalars carry no vector
current, the $Z$ boson couples to them only off-diagonally, so that the
scattering off nuclei is necessarily endothermic and the elastic channel is
absent altogether. In the pure doublet limit, however, obtaining the correct relic density together with the direct-detection (DD)
bounds on elastic spin-independent scattering leaves a tightly
constrained parameter space. A singlet admixture relaxes
these tensions considerably. The singlet--doublet mixing angle dilutes both the
gauge annihilation cross section and the off-diagonal $Z$ coupling, so that the
relic abundance and the inelastic rate can be adjusted opening a viable region of DM masses and couplings in which the LZ data can be fitted
together with all other constraints.

An interesting consequence of the framework is that the singlet--doublet mixing angle is
not a free parameter of the dark sector rather it is generated by the soft
$\mathcal{Z}_4$-breaking coupling, which is at the same time the unique source of
the Dirac neutrino mass. The strength of the
inelastic interaction and the scale of the neutrino mass are controlled by one
and the same parameter, and both vanish in the limit in which the
$\mathcal{Z}_4$ symmetry is restored. The mass splitting itself, by contrast, is
governed by the quartic coupling of the doublet to the Higgs, and remains an
independent handle with which the recoil energy of the event can be
reproduced. The interplay of these two ingredients, one tied to the neutrino
mass and one not, is what allows the framework to be predictive without being
over-constrained.

These requirements do not act independently. The relic abundance and the LZ event
both depend on the singlet-doublet mixing, which in turn fixes the neutrino mass,
while the elastic DD limits, indirect searches, and the Higgs and
electroweak precision data restrict the remaining couplings of the dark sector.
Imposed together with the neutrino oscillation data, they leave a narrow and
strongly correlated region of parameter space, with the DM mass between a few
hundred GeV and a TeV and a mass splitting of a few hundred keV. The framework is
therefore predictive rather than merely accommodating: the same measurements that
probe the DM also constrain the parameters responsible for the neutrino mass.

The rest of the paper is organized as follows. In Sec.~\ref{sec:Framework} we
introduce the model, the symmetry structure and the charge assignment, and derive
the scalar spectrum together with the one-loop Dirac neutrino mass.
Section~\ref{sec:dm} deals with the DM phenomenology, including the relic
abundance and direct and indirect detection. In Sec.~\ref{sec:lz} we compute the inelastic
recoil spectrum, confront it with the LZ event, and delineate the allowed region
of parameter space. We summarize our conclusions in Sec.~\ref{sec:conclusions}.

\section{Theoretical Framework}\label{sec:Framework}
\begin{table}[h]
\renewcommand*{\arraystretch}{1.8}
\centering
\begin{tabular}{|c|c|c|c|c|}
\hline
& Fields & $\mathcal{G}_{\rm SM}$ & $\mathcal{Z}_2$ & $\mathcal{Z}_4$ \\
\hline \hline

\multirow{3}{*}{Leptons}
& $\ell_L$ & $(\mathbf{1},\mathbf{2},-1/2)$ & $+$ & $i$ \\
& $e_R$ & $(\mathbf{1},\mathbf{1},-1)$ & $+$ & $i$ \\
& $\nu_R$ & $(\mathbf{1},\mathbf{1},0)$ & $+$ &  $-i$ \\
\hline
Fermions
& $\psi_{L},\psi_{R}$ & $(\mathbf{1},\mathbf{1},0)$ & $-, -$ & $i, i$ \\
\hline
Scalars
& $\Phi$ & $(\mathbf{1},\mathbf{2},1/2)$ & $+$ & $1$ \\
\hline
\multirow{2}{*}{Dark scalars}
& $\eta$ & $(\mathbf{1},\mathbf{2},1/2)$ & $-$ & $1$ \\
& $\chi$ & $(\mathbf{1},\mathbf{1},0)$ & $-$ & $-1$ \\
\hline
\end{tabular}
\caption{\footnotesize SM, $\mathcal{Z}_2$ and $\mathcal{Z}_4$ charge assignments of the SM and BSM fields.}
\label{tab:Z4_charge}
\end{table}

To construct a Dirac Scotogenic model that generates small Dirac neutrino masses at the one-loop level and also provides a stable DM, we extend the SM with new particles and incorporate new additional symmetries. 
To achieve this, we introduce $\mathcal{Z}_2$ and $\mathcal{Z}_4$ discrete groups on top of the SM gauge group. 
Combinations of these new symmetries ensure the Diracness of neutrinos and also take care of the stability of  DM.
In Tables~\ref{tab:Z4_charge}, we present the particle content and transformation properties of the fields under the SM,  $\mathcal{Z}_2$ and $\mathcal{Z}_4$ symmetries in our Dirac neutrino mass models.

To generate massive Dirac neutrinos, we introduce three right-handed (RH) neutrino fields $\nu_R$ into the particle content of the SM. We assign all SM fields, along with $\nu_R$, even $\mathcal{Z}_2$ charges, while the BSM fermions $\psi_{L,R}$ and the dark scalars $\eta$ and $\chi$ have odd $\mathcal{Z}_2$ charges. The stability of the DM candidate is ensured by the unbroken $\mathcal{Z}_2$ symmetry, under which the dark-sector fields are odd.
Under the $\mathcal{Z}_4$ symmetry, the lepton doublets $\ell_L$, charged-lepton singlets $e_R$, and RH neutrinos $\nu_R$ are assigned charges $i$, $i$, and $-i$, respectively, while the SM Higgs doublet $\Phi$ transforms trivially. Both BSM fermions $\psi_{L,R}$ are assigned charge $i$, whereas the BSM scalars $\eta$ and $\chi$ carry charges $1$ and $-1$, respectively.

In fact, charge assignments of $\mathcal{Z}_4$  are done in such a way that it forbids the tree-level Dirac neutrino mass term of the form $\bar{\ell_L}\tilde{\Phi}\nu_R$, but generates it at the one-loop level and ensures the Dirac nature of neutrinos as given by Fig.~\ref{fig:Nu-Dirac1loop}.
In addition, it also forbids the bare Majorana neutrino mass term  $\overline{\nu^c_R}\nu_R$ as well as the dimension-5 Weinberg operator $(\overline{\ell_L^c}\tilde{\Phi}^{*})(\tilde{\Phi}^{\dagger}\ell_L)$.
Furthermore, it is important to note that such $\mathcal{Z}_2 $ and $\mathcal{Z}_4$ charge assignments may lead to effective Majorana neutrino mass generation operators, written generically as,
\begin{equation}
\mathcal{O}_{\rm Maj}^{(6+6k)}
=
\frac{c_k}{\Lambda^{2+6k}}
\left(\nu_R^{T} C \nu_R\right)
\left[\chi\left(\eta^\dagger\Phi\right)\right]^{2k+1},
~~ k=0,1,2,..
\end{equation}
Such effective operators violate lepton number by two units at any order of  $k$. For $k=0$, this reduces to dimension-6 operator of the form:
\begin{equation}\label{eq:D6-MajMass}
\mathcal{O}_{\rm Maj}^{(6)}
=
\frac{c_0}{\Lambda^{2}}
\left(\nu_R^{T} C \nu_R\right)
\chi\left(\eta^\dagger\Phi\right).
\end{equation}
From Eq.~\ref{eq:D6-MajMass}, it can be seen that to have Majorana neutrino mass, all the scalar fields must take a vacuum expectation value (VEV).
However, in our formalism, only the SM Higgs doublet $\Phi$ takes a VEV, whereas the fields $\chi$ and $\eta$ are exempt from taking a VEV,  as addressed in a subsequent section.  This ensures that this formalism forbids a Majorana neutrino mass at any order.

The Yukawa interactions responsible for Dirac neutrino mass generation are as follows:
\begin{eqnarray}
\mathcal{L} &\supset& - y_L\overline{\ell _L}\tilde{\eta}\psi_{R}-y_R\overline{\nu_{R}}\chi\psi_{L}- m_\psi \, \overline{\psi_L}\psi_R+{\rm H.c.},
\end{eqnarray}

 The most general scalar potential allowed by the symmetries of our models is given as
 
\begin{eqnarray}
V &=& \mu_\Phi^2 (\Phi^\dagger\Phi) + \mu_\eta^2 (\eta^\dagger\eta)
   + \frac{1}{2}\mu_\chi^2 \chi^2
   + \lambda_\Phi (\Phi^\dagger\Phi)^2 
    \nonumber\\
  &+& \lambda_\eta (\eta^\dagger\eta)^2+\frac{\lambda_\chi}{4}\chi^4+ \lambda_3 (\Phi^\dagger\Phi)(\eta^\dagger\eta)
   + \lambda_4 (\Phi^\dagger\eta)(\eta^\dagger\Phi)\nonumber\\
  &+& \frac{\lambda_5}{2}\left[(\Phi^\dagger\eta)^2 + {\rm H.c.}\right]
   \nonumber+ \frac{\lambda_{\Phi\chi}}{2}(\Phi^\dagger\Phi)\chi^2
   \nonumber\\&+& \frac{\lambda_{\eta\chi}}{2}(\eta^\dagger\eta)\chi^2
   + \kappa\,\chi\left[(\eta^\dagger\Phi) + {\rm H.c.}\right] .
\label{eq:potential}
\end{eqnarray}
Note that here the ``$\kappa\,\chi(\eta^\dagger\Phi)$" terms softly break the $\mathcal{Z}_4$ symmetry, while preserves $\mathcal{Z}_2$ symmetry. Thus, plays the most significant role in generating the Dirac neutrino mass. After the electroweak symmetry breaking, $\Phi$ obtains a non-zero VEV, which leads to the mixing between $\chi$ and the real component of $\eta$, $\eta_R$. Now the squared mass matrix in $(\chi~~\eta_R)^T$ basis can be written as
\begin{equation}
\mathcal{M}^2=\begin{pmatrix}
    \mu_\chi^2+\frac{1}{2}\lambda_{\Phi\chi} v_h^2&& \kappa v_h \\
    \kappa v_h && \mu_\eta^2+\frac{(\lambda_3+\lambda_4+\lambda_5)}{2}v_h^2
\end{pmatrix}.
\end{equation}
Diagonalizing the above squared mass matrix, we obtain two mass eigenstates $\chi_1$ and $\chi_2$ with masses $m_{\chi_1}$ and $m_{\chi_2}$. The state $\chi_1$ is dominantly composed of the singlet scalar $\chi$, whereas the state $\chi_2$ is dominantly composed of the doublet scalar $\eta_R$. The mixing angle is given as
\begin{eqnarray}
\sin2\theta\simeq \frac{2\kappa v_h}{m_{\chi_2}^2-m_{\chi_1}^2}\,.
\end{eqnarray}
This mixing induces the tiny Dirac neutrino mass as shown in Fig. \ref{fig:Nu-Dirac1loop}. The neutrino mass can be calculated to be
\begin{eqnarray}
m_\nu&=&\frac{y_L \,y_R \, \sin2\theta \, m_{\psi}}{32\sqrt{2}\pi^2}  \left( \mathcal{F}\left(\frac{m_{\chi_1}^2}{m_\psi^2}\right)-\mathcal{F}\left(\frac{m_{\chi_2}^2}{m_\psi^2}\right) \right)\,,\nonumber\\
\end{eqnarray}
where the loop factor is given as $\mathcal{F}(a^2/b^2)=a^2/(a^2-b^2)$.

\begin{figure*}[t]
\centering
\includegraphics[width=0.7\linewidth]{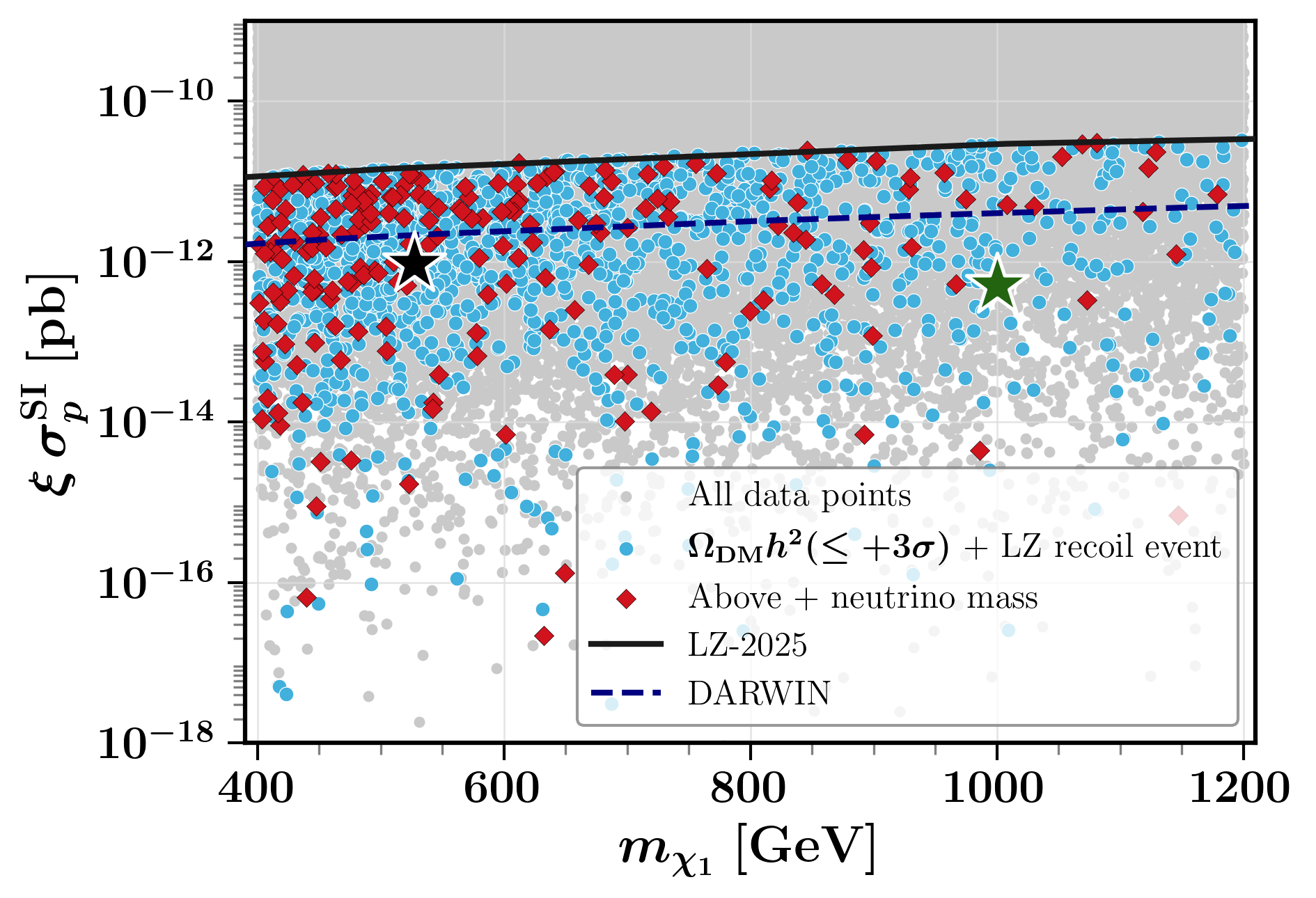}
\caption{\footnotesize The spin-independent DM-proton scattering cross section, weighted by the DM relic fraction, $\xi\,\sigma_p^{\mathrm{SI}}$, is shown as a function of the DM mass $m_{\chi_1}$. Grey points represent the full parameter scan, while the blue points satisfy the relic-density constraint within $3\sigma$ \cite{Planck:2018vyg} and reproduce the LZ recoil event with $N_{\mathrm{SR}} = 1$. In addition, the dark-red points correspond to parameter choices yielding neutrino masses of the correct order of magnitude. The solid black curve denotes the LZ-2025 exclusion limit, while the dark-blue curve shows the projected sensitivity of the future DARWIN experiment~\cite{DARWIN:2016hyl}. The black and dark green colored $\bigstar$ markers denote BP-1 and BP-2, respectively, as listed in Table~\ref{tab:BPs}.}
\label{fig:fig1}
\end{figure*}

\section{Dark matter Phenomenology}\label{sec:dm}

A structural feature of this construction turns out to be decisive for the
phenomenology. Since the singlet scalar $\chi$ remains real under the chosen symmetry and since the quartic
$\lambda_5 (\Phi^\dagger \eta)^2$ is allowed by the same symmetry, the neutral
component of $\eta$ splits into two real states of opposite CP. After electroweak
symmetry breaking, the neutral dark sector therefore consists of three real
scalars: the CP-even states $\chi_1$ and $\chi_2$, which are admixtures of $\chi$
and $\eta_R$ with a mixing angle $\theta$ generated entirely by $\kappa$, and the
CP-odd state $ \eta_I$. A real scalar has an identically vanishing vector
current, so the $Z$ boson cannot couple diagonally to any of them. Its only
coupling is the off-diagonal one between a CP-even state and $\eta_I$, arising from the doublet kinetic term as
\begin{equation}
\mathcal{L}_Z \supset -\frac{g}{2c_W} Z\mu \left[ \cos\theta \bigl(\chi_2 \overleftrightarrow{\partial^\mu} \eta_I\bigr) - \sin\theta \bigl(\chi_1 \overleftrightarrow{\partial^\mu} \eta_I\bigr) \right],\
\end{equation}
with $X \overleftrightarrow{\partial^\mu} Y \equiv X \partial^\mu Y - Y \partial^\mu X$. The
requirement that the nucleon coupling be off-diagonal, which in generic inelastic
DM constructions is an assumption, is here an exact consequence of the symmetry
that makes the neutrino Dirac. In this scenario, elastic $Z$-mediated scattering is absent, not
simply suppressed.

It is natural to  first  ask whether the Higgs portal alone can account for the LZ
event, since $\kappa$ generates an off-diagonal vertex $h\chi_1\chi_2$ alongside
the splitting. We find that it is unable to do so. The same mixing that produces the
off-diagonal vertex necessarily produces diagonal $h\chi_1\chi_1$ couplings, and
imposing the elastic spin-independent limits from the latest LZ analysis on the
resulting cross section drives the mixing and the portal quartics down to values
for which the inelastic rate falls well below one event in the
$2.84\,\mathrm{tonne\!-\!yr}$ exposure for the local halo flux. The scalar portal
is thus unable to reconcile the two requirements simultaneously.

The resolution is that the dominant interaction is not scalar but vector
mediated. The DM candidate $\chi_1$ up-scatters off a nucleus into the CP-odd
state, $\chi_1 N \to \eta_I  N$, through $Z$ exchange with a coupling
$\tfrac{g}{2\cos\theta_W}\sin\theta$ inherited from the $\eta_R$ admixture of
$\chi_1$. The rate is governed by $\sin^2\theta$ and is of weak strength. It is several orders of magnitude larger than what can be obtained from the Higgs portal. Therefore, a single event in the LZ exposure can be obtained for $\delta_{RI} \equiv m_{\eta_I} - m_{\chi_1}$ of a few hundred keV. At the same time, there is no conflict with the elastic limits, as the diagonal coupling vanishes identically. The relevant splitting for the LZ event is therefore $\delta_{RI}$
rather than the CP-even splitting, and the relevant coupling is $\sin\theta$.

In Fig.~\ref{fig:fig1}, we demonstrate the SI $\chi_1$-proton cross section, weighted by the relic fraction $\xi\equiv\Omega_{\rm DM}h^2/\Omega^{\rm Planck}_{\rm DM}h^2$, over the scan range $m_{\chi_1} \in [400, 1200]\,\mathrm{GeV}$, $\delta_1 \equiv m_{\chi_2} - m_{\chi_1} \in [10, 100]\,\mathrm{GeV}$, $\theta \in [0, \pi/2]$, the independent quartic couplings sampled logarithmically over $[10^{-3}, 3]$ and the Yukawa couplings are scanned over the range $[10^{-5},10^{-3}]$ assuming, $y_L=y_R$ for simplicity. Here we fix $m_\psi=10^5$ GeV. All scanned points satisfy the theoretical and phenomenological requirements, including vacuum stability, boundedness from below, perturbative unitarity, electroweak precision constraints, Higgs-sector constraints, and limits from charged lepton flavor violating (cLFV) processes \cite{MEGII:2023ltw} and muon $g-2$ \cite{Boccaletti:2024guq, Muong-2:2025xyk,Aliberti:2025beg}. The details of $(g-2)_\mu$ and cLFV are given in Appendix \ref{app:g-2}. We implemented the model in \texttt{FeynRules} \cite{Alloul:2013bka}, and the relic abundance and DD cross sections were computed with \texttt{micrOMEGAs} \cite{Belanger:2026asz,Alguero:2023zol,Belanger:2018ccd}. Whereas, the Higgs sector constraints were checked through \texttt{HiggsBounds} \cite{Bechtle:2020pkv} and \texttt{HiggsSignals} \cite{Bechtle:2020uwn}. In Fig.~\ref{fig:fig1}, blue colored points satisfy the relic-density bound within $+3\sigma$ together with the LZ recoil requirement $N_{\mathrm{SR}} \simeq 1$ (defined below). The dark-red diamonds are the subset of these model points that additionally reproduce light neutrino masses of the correct order of magnitude. The solid black curve is the LZ-2025 SI exclusion \cite{LZ:2024zvo}, and the dark blue colored dashed curve is the projected DARWIN sensitivity \cite{DARWIN:2016hyl}. As anticipated above, the elastic rate here is generated entirely through the Higgs portal, since the $Z$ coupling is exactly off-diagonal, and a substantial population survives the SI limit, spanning several decades in $\xi\,\sigma_p^{\mathrm{SI}}$, as shown.

\begin{table*}[t]
\centering
\begin{tabular}{| c|c| c| c| c| c| c| c| c| c| c|} 
 \hline
 BP& $m_{\chi_1}$ (GeV) &  $m_{\chi_2}$(GeV) &  $\delta_{RI}$ (keV)&  $m_{\eta^+}$ (GeV) & $\sin\theta$ & $y_L,y_R$ & DM relic & DD elastic & Neutrino mass & $(g-2)_\mu\&$cLFV \\
 \hline
 BP1 & 527.19  & 541.21& 342.12 &531.04 & 0.55&$1.3\times10^{-4}$ & $\checkmark$  & $\checkmark$ & $\checkmark$ & $\checkmark$  \\
 \hline
  BP2 & 1000.00  & 1016.53& 355.79 &1007.43 & 0.32&$1.15\times10^{-4}$ & $\checkmark$  & $\checkmark$ & $\checkmark$ & $\checkmark$  \\
 \hline
\end{tabular}
\caption{\footnotesize Two benchmark points that give the correct DM relic, neutrino mass, $(g-2)_\mu$ and, cLFV. These BPs are used in LZ 248 keV event analysis in Section \ref{sec:lz}.}\label{tab:BPs}
\end{table*}

\section{LUX-ZEPLIN 248 keV recoil event}\label{sec:lz}
In this section, we study the compatibility of our scenario with the observed LZ230616 event. We calculate the expected number of signal events, taking into account the nuclear form factor, detector efficiency and recoil-energy smearing. We then perform a likelihood analysis to identify the preferred regions of the parameter space that can accommodate the observed event. The differential rate per unit detector mass is given by \cite{Tucker-Smith:2001myb}
\begin{eqnarray}
    \frac{d\mathcal{R}}{dE_R}=N_T\frac{\xi \rho_{\chi_1}}{m_{\chi_1}}\int_{v_{\rm min}}^{v_{\rm max}}dv ~vf(v)\frac{d\sigma}{dE_R},
    \label{eq:dRdER}
\end{eqnarray}
where $N_T$ is the number of target nuclei per unit detector mass,
$\rho_{\chi_1}$ denotes the local DM energy density associated with the
ground-state component, $\xi$ is the relative DM abundance, and $f(v)$ is the DM speed
distribution in the detector frame. The speed distribution $f(v)$ is given by
\begin{eqnarray}
    f(v)=\frac{v}{\sqrt{\pi}v_ev_0}e^{-\frac{v_e^2+v^2}{v_0^2}}\left( e^{\frac{2vv_e}{v_0^2}}- e^{-\frac{2vv_e}{v_0^2}} \right),
\end{eqnarray}
with $v_0=238$ km/s \cite{Baxter:2021pqo}, and the time-averaged Earth's speed relative to the galactic rest frame is $v_e=v_\odot=v_0+12 {~\rm km/s}$. For the scalar-mediated interaction considered here, the spin-independent differential cross-section can be expressed as:
\begin{eqnarray}
  \frac{d\sigma}{dE_R}=\frac{m_N}{2v^2}\frac{\sigma_n}{\mu_n^2}\frac{(f_pZ+f_n(A-Z))^2}{f_n^2}F^2(E_R),
\end{eqnarray}
with the Helm form factor:
\begin{eqnarray}
    F^2(E_R)=\left( \frac{3j_1(qr_0)}{qr_0} \right)^2e^{-s^2q^2},
\end{eqnarray}
where, $q=\sqrt{2m_NE_R}$, $s=1$ fm, $r_0=\sqrt{r^2-5s^2}$, $r=1.2A^{1/3}$ fm\footnote{Note that we also cross-checked and verified our results using publicly available nuclear response functions \cite{WIMpy-code,Jeong:2021bpl}.}. 
Restricting the integral to the galactic escape velocity $v_{\rm max}=v_{\rm esc}+v_e$ (using $v_{\rm esc}=544~{\rm km/s}$  \cite{Baxter:2021pqo}), we obtain:
\begin{eqnarray}
 \frac{d\mathcal{R}}{dE_R}&=&\frac{N_Tm_N\rho_\chi}{4v_0m_{\chi_1}}\frac{\sigma_n}{\mu_n^2}\frac{(f_pZ+f_n(A-Z))^2}{f_n^2}F^2(E_R) \nonumber\\&& \times\frac{1}{\eta}\bigg({\rm erf}[x_{\rm min}+\eta]-{\rm erf}[x_{\rm min}-\eta]\nonumber\\&&-{\rm erf}[x_{\rm max}+\eta]+{\rm erf}[x_{\rm max}-\eta] \bigg), \label{eq:diff_rate} 
\end{eqnarray}
where, $x_{\rm min}=v_{\rm min}/v_0$, $\eta=v_e/v_0$ and $x_{\rm max}=v_{\rm max}/v_0$.

We define the parameter vector
\begin{equation}
    \Theta =
    \left\{
        m_{\chi_1},\,
        \delta_{RI},\,
        \sin\theta
    \right\}.
    \label{eq:LZ_parameters}
\end{equation}
For a given point in this parameter space, the predicted number of
events per unit recoil energy is
\begin{equation}
    \frac{dN(\Theta)}{dE_R}
    =
    \mathcal{E}_{\rm LZ}\,
    \epsilon(E_R)\,
    \frac{d\mathcal R(\Theta)}{dE_R},
    \label{eq:dNdER_LZ}
\end{equation}
where $d\mathcal R/dE_R$ is the differential recoil rate derived in
Eq.~\ref{eq:dRdER}. In Eq.~(\ref{eq:dNdER_LZ}), $\epsilon(E_R)$ denotes the detector efficiency and $\mathcal{E}_{\rm LZ}=2.84\,\mathrm{tonne\!-\!yr}$ is the exposure used in the high-energy recoil LZ analysis \cite{LZ:2026axp}.
We convolve the predicted rate with LZ’s region of interest (ROI) efficiency and with a Gaussian energy response of width $\sigma_E(E_R) = \left[ \frac{(23\text{ keV})^2 E_R}{248\text{ keV}} + (0.093 E_R)^2 \right]^{1/2}$ \cite{Rodd:2026tyn}. The expected total number of signal events in the analysis window is
then,
\begin{equation}
    N_{\rm SR}(\Theta)
    =
    \int_{E_R^{\rm min}}^{E_R^{\rm max}}
    dE_R\,
    \frac{dN(\Theta)}{dE_R},
    \label{eq:Ntot_LZ}
\end{equation}
where, $E^{min}_R=5.4,\mathrm{keV}$ and $E^{max}_R=269.9,\mathrm{keV}$.

For the single event of interest, the extended unbinned likelihood can
be written as \cite{Barlow:1990vc}
\begin{equation}
    \mathcal L(\Theta)
    =
    e^{-N_{\rm SR}(\Theta)}
    \prod_{i=1}^{n_0}
    \left.
    \frac{dN(\Theta)}{dE_R}
    \right|_{E_R=E_i},
    \label{eq:LZ_extended_likelihood}
\end{equation}
where, $n_0=1$ and $E_i = 248 \pm 23 ~(\rm stat) \pm23 ~(sys)$ keV.

We construct the negative log-likelihood as \cite{Barlow:1990vc, Wu:2026nhi, Su:2026rwz}:
\begin{equation}
    -\ln\mathcal L(\Theta)
    =
    N_{\rm SR}(\Theta)
    -
    \left.
    \ln\left[
    \frac{dN(\Theta)}{dE_R}
    \right]\right|_{E_R=248~{\rm keV}},
    \label{eq:minuslogL}
\end{equation}
and minimize it to extract the best-fit (BF) model parameters, $\hat{\Theta} \equiv (\hat{m}_{\chi_1}, \hat{\delta}_{RI}, \hat{\sigma}_n)$, yielding the maximum likelihood:
\begin{equation}
    -\ln\mathcal L_{\rm max}
    =
    -\ln\mathcal L(\widehat{\Theta}).
    \label{eq:Lmax}
\end{equation}
We define the test statistic
\begin{equation}
    {\rm TS}(\Theta)
    =
    2
    \left[
        -\ln\mathcal L(\Theta)
        +\ln\mathcal L_{\rm max}
    \right].
    \label{eq:TS_LZ}
\end{equation}
\begin{figure*}[!htpb]
    \centering
    \includegraphics[height=5.4cm,width=7.7cm]{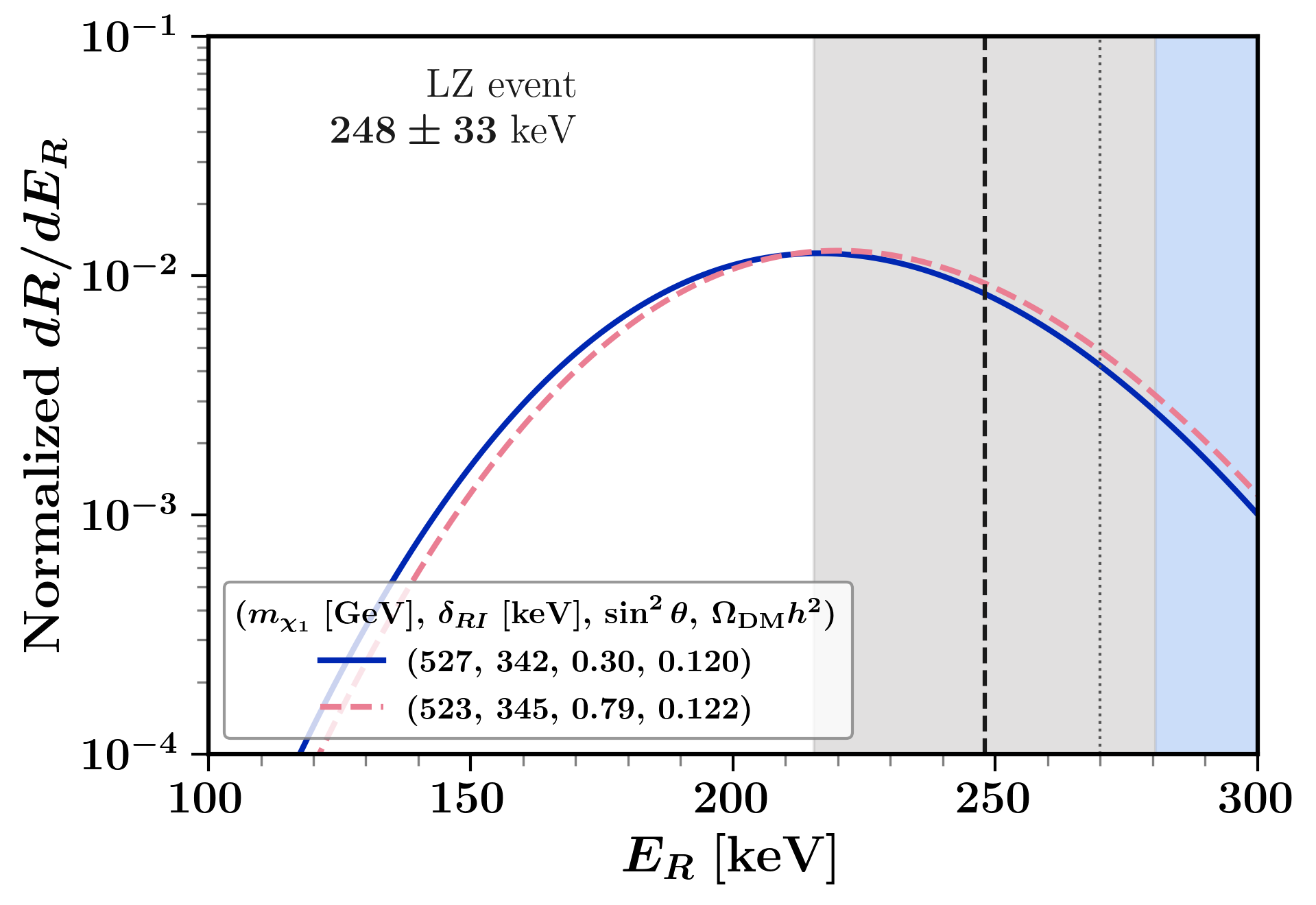}
    \hspace{0.2cm}
    \includegraphics[height=5.4cm,width=7.7cm]{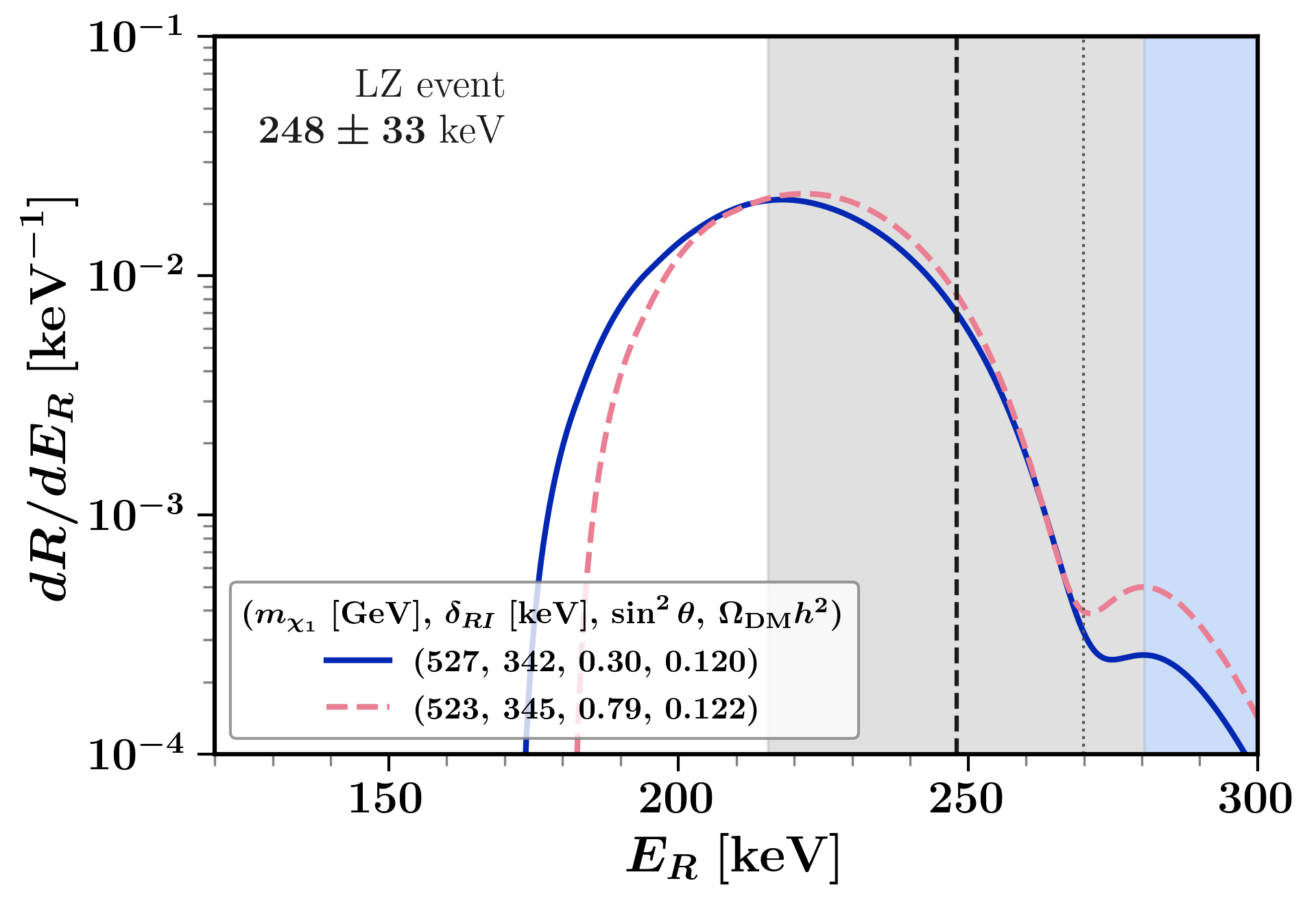}

    \includegraphics[height=5.4cm,width=7.7cm]{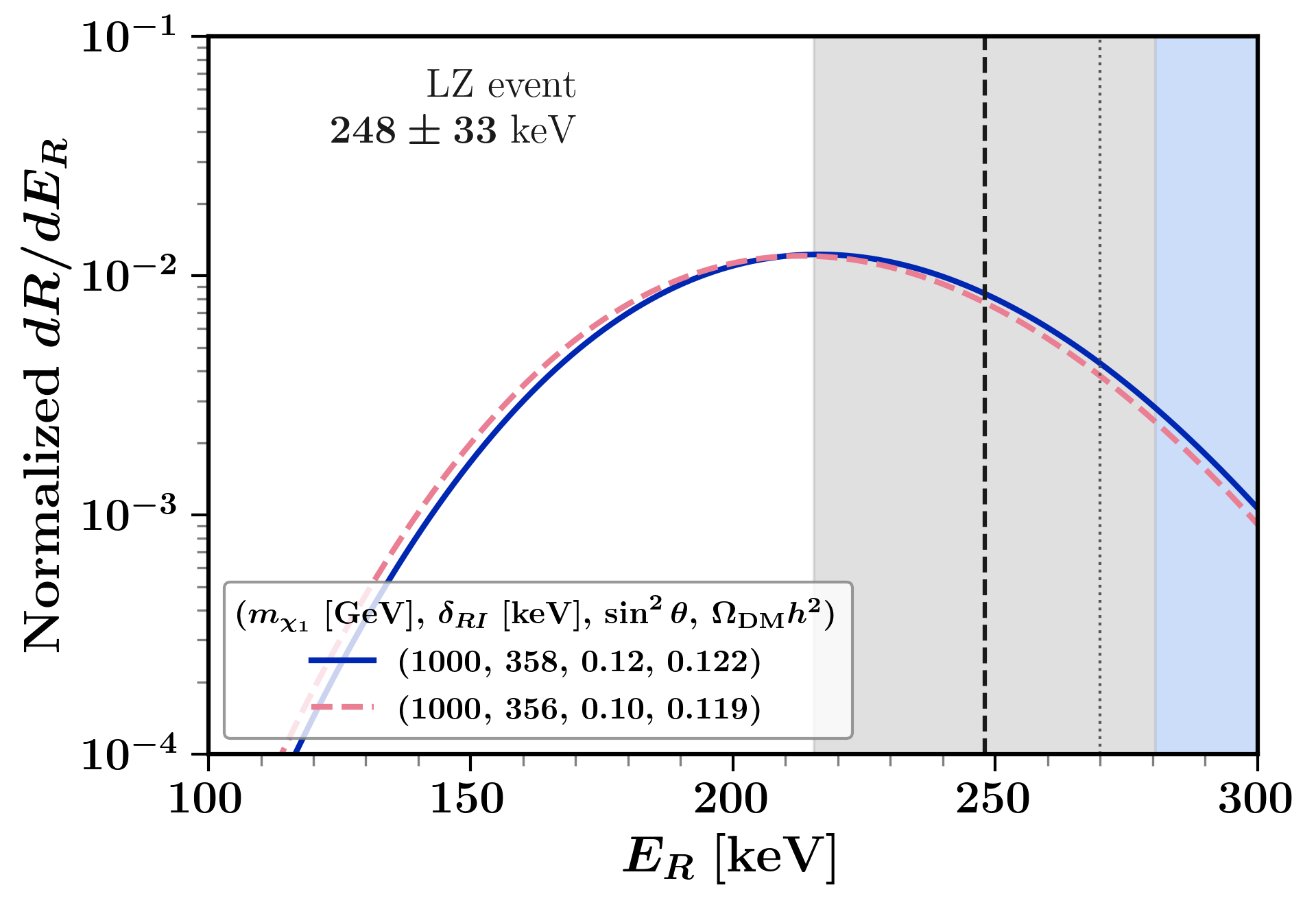}
    \hspace{0.2cm}
    \includegraphics[height=5.4cm,width=7.7cm]{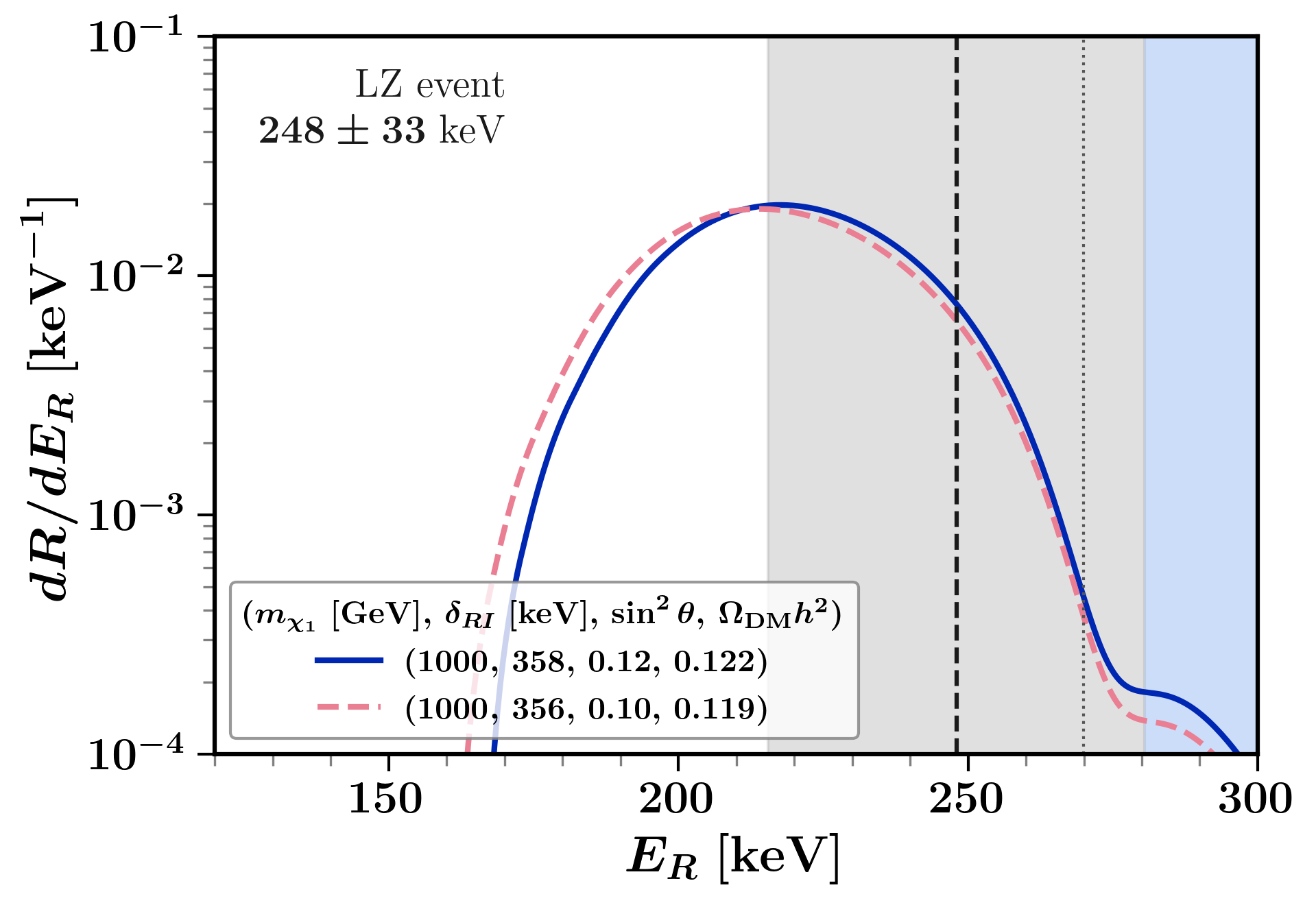}

    \vspace{-0.2cm}
    \makebox[0.48\textwidth][c]{(a)}
    \hfill
    \makebox[0.48\textwidth][c]{(b)}

    \caption{\footnotesize Differential recoil spectra, $dR/dE_R$, for representative benchmark points in the inelastic DM parameter space. The upper and lower rows correspond to $m_{\chi_1} \simeq 500\,\mathrm{GeV}$ and $m_{\chi_1} = 1\,\mathrm{TeV}$, respectively. Panel (a) shows the normalized predicted spectra weighted by the LZ detection efficiency ($\epsilon(E_R)$) and convolved with the Gaussian detector response, while panel (b) displays the corresponding spectra without Gaussian energy smearing. The vertical dashed line denotes the central recoil energy, $E_R = 248\,\mathrm{keV}$, of the LZ event, and the grey shaded band indicates the $248 \pm 33\,\mathrm{keV}$ recoil-energy interval. The blue shaded region represents the high-energy sideband above the LZ signal region of interest.}
    \label{fig:3}
\end{figure*}

The likelihood in Eq.~(\ref{eq:minuslogL}) therefore incorporates both the overall normalization and the spectral preference associated with the observed event. The first term, $N_{\text{SR}}(\Theta)$, suppresses parameter points that predict an excessive number of events over the full LZ recoil window, while the second term favors parameter configurations for which the differential recoil rate is appreciable at the observed energy, $E_R = 248\text{ keV}$. Consequently, the fit is sensitive not only to the magnitude of the inelastic scattering rate but also to the kinematic shape of the recoil spectrum. In particular, in endothermic scattering, the mass splitting $\delta_{RI}$ controls the portion of the high-velocity tail contributing to a given recoil energy.

\begin{figure*}[!htpb]
\centering
\includegraphics[width=0.45\linewidth]{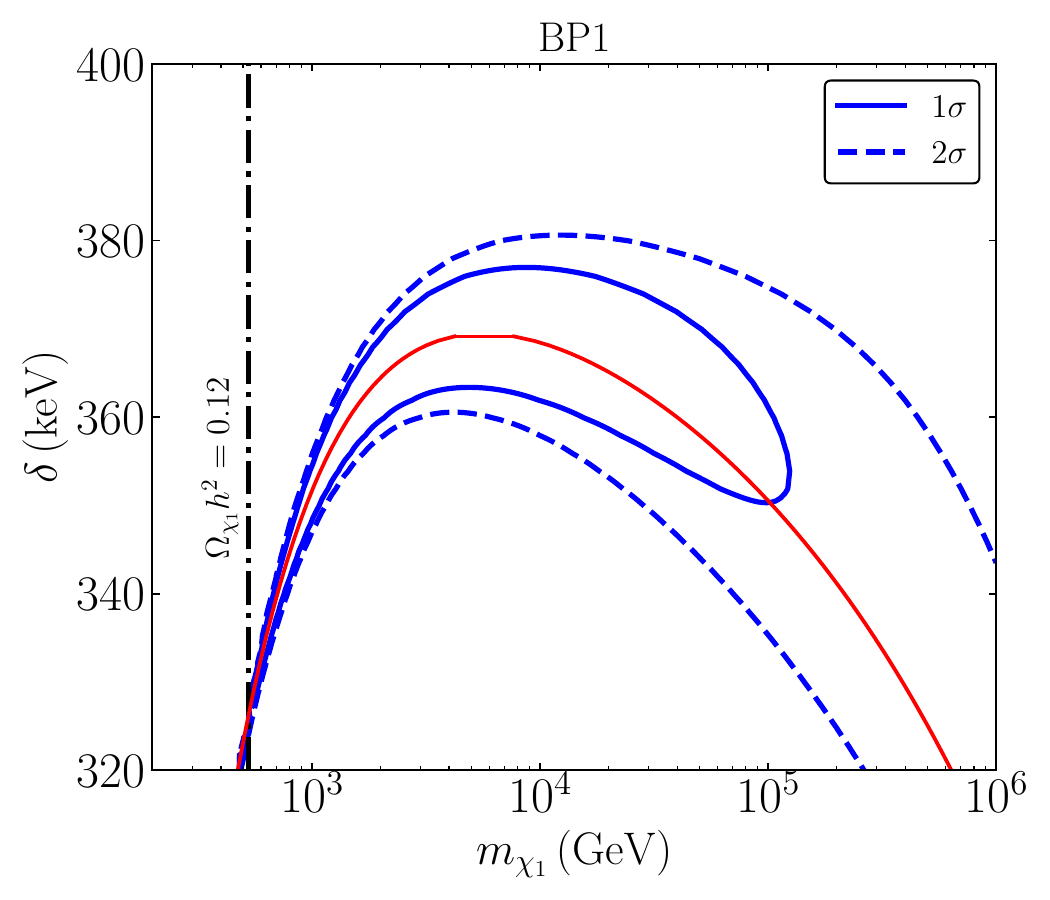}
\includegraphics[width=0.45\linewidth]{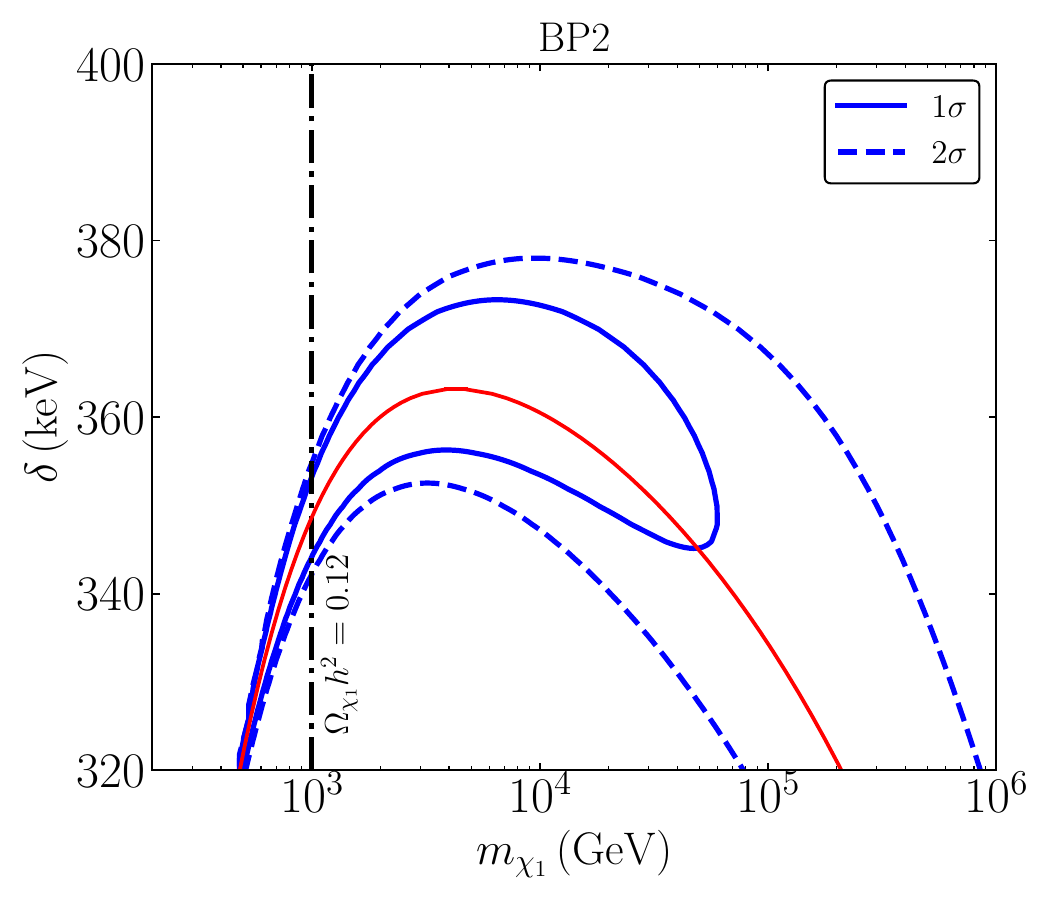}
    \vspace{-0.2cm}
    \makebox[0.48\textwidth][c]{(a)}
    \hfill
    \makebox[0.48\textwidth][c]{(b)}
\caption{\footnotesize (a) $1\sigma$ and $2\sigma$ contours accommodating the LZ230616 event in the $\delta -m_{\chi_1}$
plane for $\sigma^{\rm inel}_N=7\times10^{-40}{~\rm cm^2}$. The red contour represents $N_{\rm SR}=1$. The black dashed-dotted line corresponds to the correct DM relic point $\{m_{\chi_1}=527.19{\rm~ GeV}\}$. (b) Same as in panel (a) but for $\sigma^{\rm inel}_N=2.31\times10^{-40}{~\rm cm^2}$. The red contour represents $N_{\rm SR}=1$, and the black dashed-dotted line corresponds to the correct DM relic point $\{m_{\chi_1}=1{\rm~ TeV}\}$.}
\label{fig:lzBP1}
\end{figure*}

The recoil spectra corresponding to two sets of representative benchmark points (BPs), selected by the likelihood analysis, are illustrated in Fig.~\ref{fig:3}. The upper and lower panels correspond to sub-TeV DM mass with $m_{\chi_1} \simeq 500\text{ GeV}$ and fixed DM mass $m_{\chi_1} = 1~{\rm TeV}$, respectively. In panel (a), the predicted recoil rates are weighted by the LZ detection efficiency $\epsilon(E_R)$ and convolved with the Gaussian detector response, characterized by $\sigma_E (E_R)$, and subsequently normalized to unit integral during the analysis window. The abscissa is, therefore, the reconstructed energy, directly comparable to the observed event. Each spectrum rises through the ROI, peaks close to $248\,\mathrm{keV}$, and falls off smoothly beyond it. This behavior is a direct consequence of endothermic kinematics:  nonzero splitting $\delta_{RI}$ introduces an interior minimum in $v_{\min}(E_R)$, causing the recoil spectrum to be concentrated around the corresponding characteristic recoil energy rather than decreasing monotonically with $E_R$. The Gaussian energy response further smooths the sharp kinematic features. 

For the two sets of BPs shown in panel (a), between $60\%$ and $70\%$ of the total in-window rate falls within the signal region $E_R \in [200, 269.9]\,\mathrm{keV}$, which is consistent with a spectrum that is weighted toward high recoil energy, as required to accommodate the event. Within a given mass pair, the two curves are nearly degenerate in shape across the region of interest. They agree to within a few percent for $E_R \gtrsim 200\,\mathrm{keV}$, and differ appreciably only below $\sim 150\,\mathrm{keV}$, where the rate has already fallen to $\lesssim 20\%$ of its peak value. This near-degeneracy is exactly what one expects, since $\sin^2\theta$ enters $d\sigma/dE_R$ only as an overall multiplicative constant and therefore cannot shift where the spectrum sits. It can only rescale its height. 

However, the singlet-doublet admixture provides an important degree of freedom absent in the pure electroweak doublet scenario. Since the $Z$-mediated inelastic scattering rate scales as $\sigma_N^{\mathrm{inel}} \propto \sin^2\theta$, the singlet component allows the interaction strength to be varied independently of the inelastic mass splitting, thereby adjusting the signal normalization while retaining the required high-recoil kinematics. In contrast, for a pure electroweak doublet, the interaction strength is fixed by its gauge quantum numbers, leaving substantially less freedom to simultaneously satisfy the relic-density, SI DD, and LZ event-rate requirements.

Panel (b) shows the same BPs with $\epsilon_{\mathrm{LZ}}(E_R)$ still applied, but without the Gaussian smearing of panel (a). Each spectrum now ends at a sharp endothermic edge, $E_R^-$, below which the rate vanishes identically rather than simply becoming small. It is the rise of this edge $E_R^-$ with $\delta$ that lets the $248\,\mathrm{keV}$ feature select $\delta = \mathcal{O}(340-360)\,\mathrm{keV}$ in the first place. The accessible window $[E_R^-, E_R^+]$ has to include the event. There is also a small residual structure near $E_R \sim 270-290\,\mathrm{keV}$ (a shallow dip followed by a partial rise) visible in both mass pairs. This reflects the diffraction minimum of the Helm nuclear form factor for xenon at the momentum transfers relevant here, $q \simeq 1.1-1.3\,\mathrm{fm}^{-1}$ across the analysis window (with $\epsilon_{\mathrm{LZ}}(E_R)$ applied). The corresponding $\Delta\chi^2$ profiles for these benchmark points are given in Appendix~\ref{app:chisq}.

In the panel (a) of Fig. \ref{fig:lzBP1}, the resulting $1\sigma$ and $2\sigma$ confidence regions accommodating the LZ230616 event are shown in the $(m_{\chi_1},\delta)$ plane for $\sigma^{\rm inel}_N=7\times10^{-40}{~\rm cm^2}$. The details of the benchmark point is provided in Table \ref{tab:BPs}. The solid and dashed blue contours correspond to the $1\sigma$ and $2\sigma$ regions, respectively, while the red contour represents the parameter space for which the expected number of signal events is exactly one. The black dashed-dotted vertical line indicates the DM mass that gives the correct DM relic abundance, corresponding to $m_{\chi_1}=527.19{\rm~ GeV}$, satisfying neutrino mass, $(g-2)_\mu$, and cLFV. The correct relic point lies within the preferred region that accommodates the observed LZ230616 event. The characteristic shape of the confidence regions can be understood as follows. For smaller values of $m_{\chi_1}$, the allowed mass splitting $\delta$ increases with increasing $m_{\chi_1}$, reaching a maximum around $m_{\chi_1}\sim 10^4~{\rm GeV}$. As $m_{\chi_1}$ increases, a larger mass splitting can be accommodated while still allowing the required recoil energy to be produced by the high velocity DM particles. However, for $m_{\chi_1}\gg m_N$, the reduced mass approaches the nucleus mass and the kinematic dependence on $m_{\chi_1}$ becomes weak. At the same time, the DM number density decreases with increasing $m_{\chi_1}$, resulting in a suppression of the scattering rate. Consequently, at larger $m_{\chi_1}$, smaller values of $\delta$ are preferred to compensate for the reduction in the event rate. This leads to the characteristic turnover and the gradual decrease of $\delta$ at large $m_{\chi_1}$. The red $N_{\rm SR}=1$ contour follows a similar trend and provides the boundary corresponding to one expected signal event. The benchmark point satisfying the correct relic abundance also predicts a signal rate of the appropriate order to account for the observed LZ230616 event. In the panel (b) of Fig. \ref{fig:lzBP1}, we show another benchmark point, BP2, with $\sigma^{\rm inel}_N=2.31\times10^{-40}{\rm~ cm^2}$. For this benchmark point, the correct DM relic abundance is obtained for $m_{\chi_1}=1$ TeV. This parameter point, shown by the black dashed-dotted line, lies within the $1\sigma$ allowed region obtained from the LZ230616 event. Thus, both BP1 and BP2 simultaneously satisfy the correct DM relic abundance and provide a good fit to the observed LZ230616 event.
\section{Conclusions}\label{sec:conclusions}
We propose a formalism in which a one-loop radiative neutrino mass model naturally realizes inelastic dark matter, offering an explanation for the recent $248~\mathrm{keV}$ nuclear-recoil event observed by LUX-ZEPLIN, while simultaneously accounting for neutrino masses and their Dirac nature.
As in the SM, a renormalizable Dirac neutrino mass term is not allowed by the SM gauge symmetries. To generate neutrino masses, we introduce three right-handed neutrinos in the SM particle content. We further extend the model by introducing pairs of BSM fermions and a pair of dark scalars, consisting of one scalar doublet and one scalar singlet.
Keeping in mind our goals, we impose a $\mathcal{Z}_2 \times \mathcal{Z}_4$ symmetry, which forbids the dimension-4 Dirac neutrino mass term and prevents Majorana mass terms at all orders, including Majorana masses of both $\nu_R$ and the
vector-like fermions. 

Having said that, we introduce a soft-breaking term  $\kappa\,\chi(\eta^\dagger\Phi)$ that helps us to generate Dirac neutrino masses at the one-loop level (see Fig.~\ref{fig:Nu-Dirac1loop}). Once the SM symmetry is spontaneously broken by the SM Higgs VEV, mixing between the dark scalars is induced, leading to the desired DM candidate.
Thus, this framework simultaneously generates Dirac neutrino masses and provides a viable DM candidate that could potentially explain the latest LUX-ZEPLIN observed event. As the symmetry allows $\chi$ to be a
real scalar, and because the quartic $\lambda_5(\Phi^\dagger\eta)^2$ is permitted
by the same symmetry, the neutral component of $\eta$ splits into two real states
of opposite CP. Every neutral dark state is therefore a real scalar, and since a
real field carries no vector current, the $Z$ boson couples to them only
off-diagonally. Therefore, the elastic $Z$-mediated scattering is identically absent, and
the off-diagonal nucleon coupling that the inelastic DM normally has to
assume here is an exact consequence of the symmetry responsible for the Dirac
nature of the neutrino. We find that the Higgs portal alone cannot account
for the event, because the same mixing that generates the off-diagonal vertex
$h\chi_1\chi_2$ unavoidably generates the diagonal coupling $h\chi_1\chi_1$,
whose elastic rate is already constrained. The viable channel is the
vector-mediated up-scattering $\chi_1 N \to \eta_I N$, whose strength is
controlled by singlet-doublet mixing $\sin\theta$.

The interplay between the two sectors (i.e., the neutrino and the DM) is what makes the framework predictive. The
singlet--doublet mixing angle is not a free parameter. 
It is generated by `$\kappa$', which simultaneously generates the Dirac neutrino mass and  controls the inelastic rate. The splitting $\delta_{RI}$, on the contrary,
is controlled by the quartic couplings of the doublet and remains an independent
handle, with which the position of the recoil spectrum is fixed. Scanning the parameter space subject to theoretical and experimental
constraints, we find a viable region with $m_{\chi_1}$ between roughly
$400\,$GeV and $1.2\,$TeV and $\delta_{RI} \simeq 340$--$360\,$keV, in which the
observed relic abundance, light neutrino masses of the correct order and
$N_{\rm SR}\simeq1$ in the $2.84\,$tonne-yr exposure are reproduced
simultaneously. We summarize our noteworthy results in Figs.~\ref{fig:fig1}, \ref{fig:3}, and \ref{fig:lzBP1}.

The framework carries several testable implications. This explains the Dirac nature of neutrinos along with their tiny masses. The surviving elastic
cross sections extend over several orders of magnitude in $\xi\sigma^{\rm SI}_p$, a
substantial portion of which lies above the anticipated sensitivity, so
the scenario will be probed further even in the absence of additional
high-energy events. Since the expected number of inelastic events scales linearly
with exposure, the full LZ dataset and next-generation xenon observatories will
either accumulate a spectrum whose shape would then discriminate the endothermic
interpretation from alternatives, or exclude it. Should the event prove to be a
background fluctuation, the correlation established here between the
soft-breaking parameter, the inelastic rate and the neutrino mass remains a
generic feature of the construction and continues to constrain it.

\section*{Acknowledgment}
We thank Partha Kumar Paul for his involvement in the early stages of this
project and for his help with the analysis of the LZ event.
P. B. acknowledges the financial support received from the Indian Institute of Technology, Guwahati (IITG) as an Institute Post-Doc Fellow (IPDF), grant {IITG/AR/IPDF/2026-27/024}. S.M. acknowledges support from the IIT Goa Startup Grant
[2025/SG/SM/057].

\appendix

\section{$(g-2)_\mu$ and charge lepton flavor violation (cLFV)}\label{app:g-2}

\begin{figure}[H]
\centering
\includegraphics[width=0.8\linewidth]{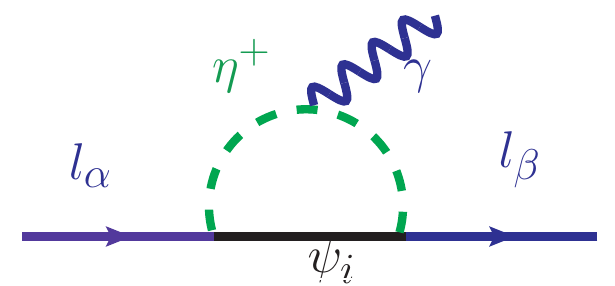}
\caption{\footnotesize Contribution to $(g-2)_\mu$ and cLFV from $\eta^+$ and $\psi_i$ in one loop.}
\label{fig:g-2}
\end{figure}

The charge component of $\eta$ and $\psi_i$ give rise to a negative contribution to the ($g-2$) of muon as shown in Fig. \ref{fig:g-2}. The contribution is estimated to be \cite{Queiroz:2014zfa}
\begin{eqnarray}
\Delta a_\mu=\sum_i -\frac{m_\mu^2}{8\pi^2 m_{\eta^+}^2}|y^{\mu i}_{L}|^2f(m_{\psi_i^2}/m_{\eta^+}^2),
\end{eqnarray}
where the loop factor is given as
\begin{eqnarray}
f(x)=\frac{1-6x+3x^2+2x^3-6x^2\log x}{12(1-x)^4}.
\end{eqnarray}

In our framework, the same diagram, Fig. \ref{fig:g-2}, also gives rise to charge lepton flavor violation (cLFV) via the decay $\mu\rightarrow e\gamma$. The branching ratio for the process can be calculated as \cite{Staub:2016dxq}
\begin{eqnarray}
{\rm Br} (\mu \rightarrow e \gamma) =\frac{3 (4\pi)^3 \alpha}{4G^2_F} (\lvert A_{e\mu}^M \rvert^2+\lvert A_{e\mu}^E \rvert^2) {\rm Br} (\mu \rightarrow e \nu_{\mu} \overline{\nu_e}).
\end{eqnarray}
where, $A_{e\mu}^{M, E}$ are the dipole form factors defined as
\begin{eqnarray}
	A_{e\mu}^M=\frac{-1}{(4\pi)^2}\sum_{i}((y^{e i}_L)^* y_L^{\mu i} I^{++}_i+(y_L^{e i})^* y_L^{\mu i} I^{+-}_i),
\end{eqnarray}
\begin{eqnarray}
	A_{e\mu}^E=\frac{-i}{(4\pi)^2}\sum_{i}(-(y_L^{e i})^* y_L^{\mu i} I^{-+}_k-(y_L^{e i})^* y_L^{\mu i} I^{--}_i),
\end{eqnarray}
where 
\begin{eqnarray}
I^{(\pm)_1(\pm)_2}_{i}=\int d^3X\frac{x(y+(\pm)_1z \frac{m_{e}}{m_{\mu}}+(\pm)_2\frac{m_{\psi_i}}{m_{\mu}})}{-xy m_{\mu}^2-xzm_e^2+(1-x)m_{\eta^+}^2+x m_{\psi_i}^2}\nonumber\\
\end{eqnarray}

\section{$\chi^2$ fitting for LZ230616.}
\label{app:chisq}
\begin{figure*}[]
    \centering
    \hspace{0.2cm}
    \includegraphics[height=5.4cm,width=7.3cm]{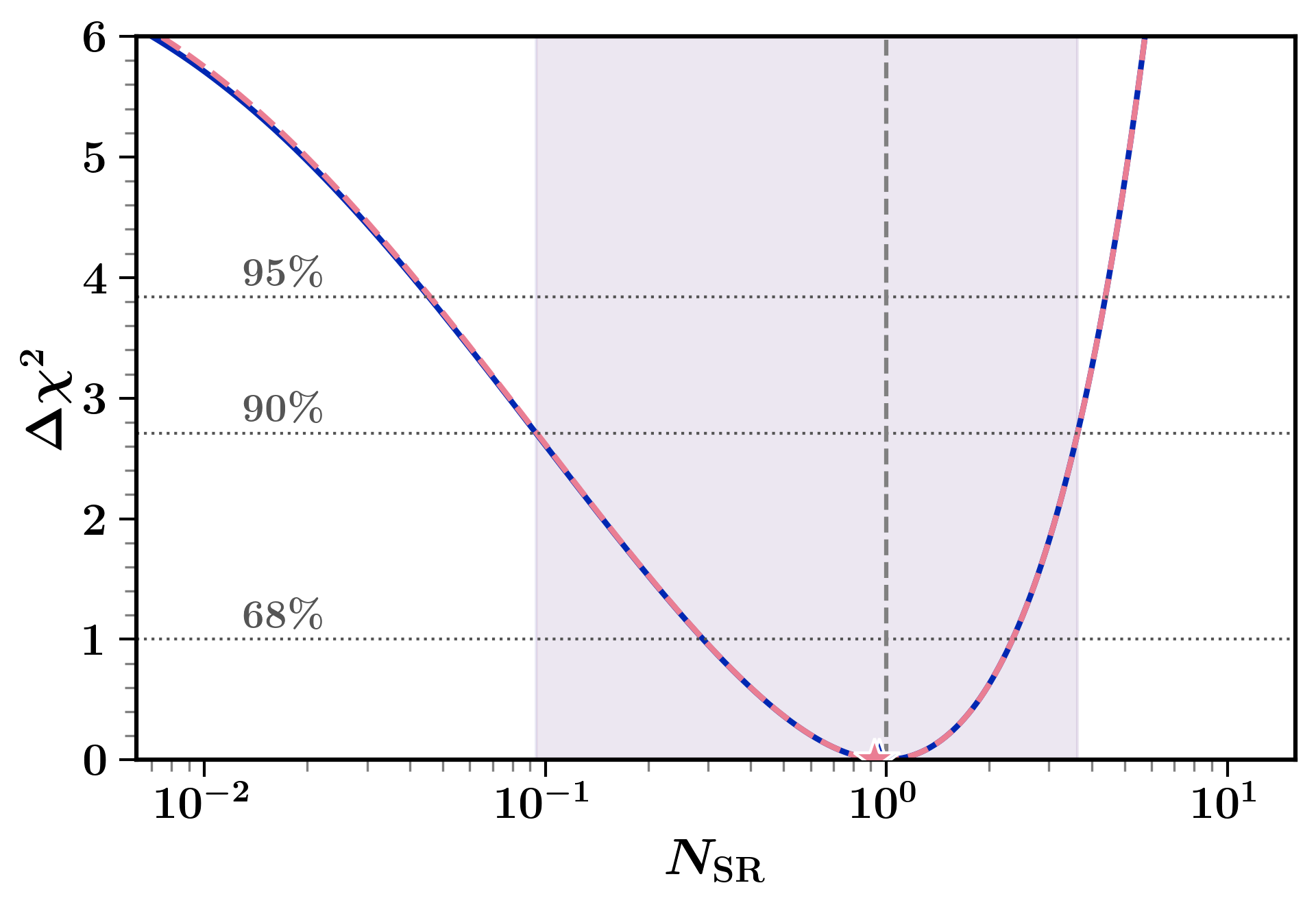}
    \hspace{0.2cm}
    \includegraphics[height=5.4cm,width=7.3cm]{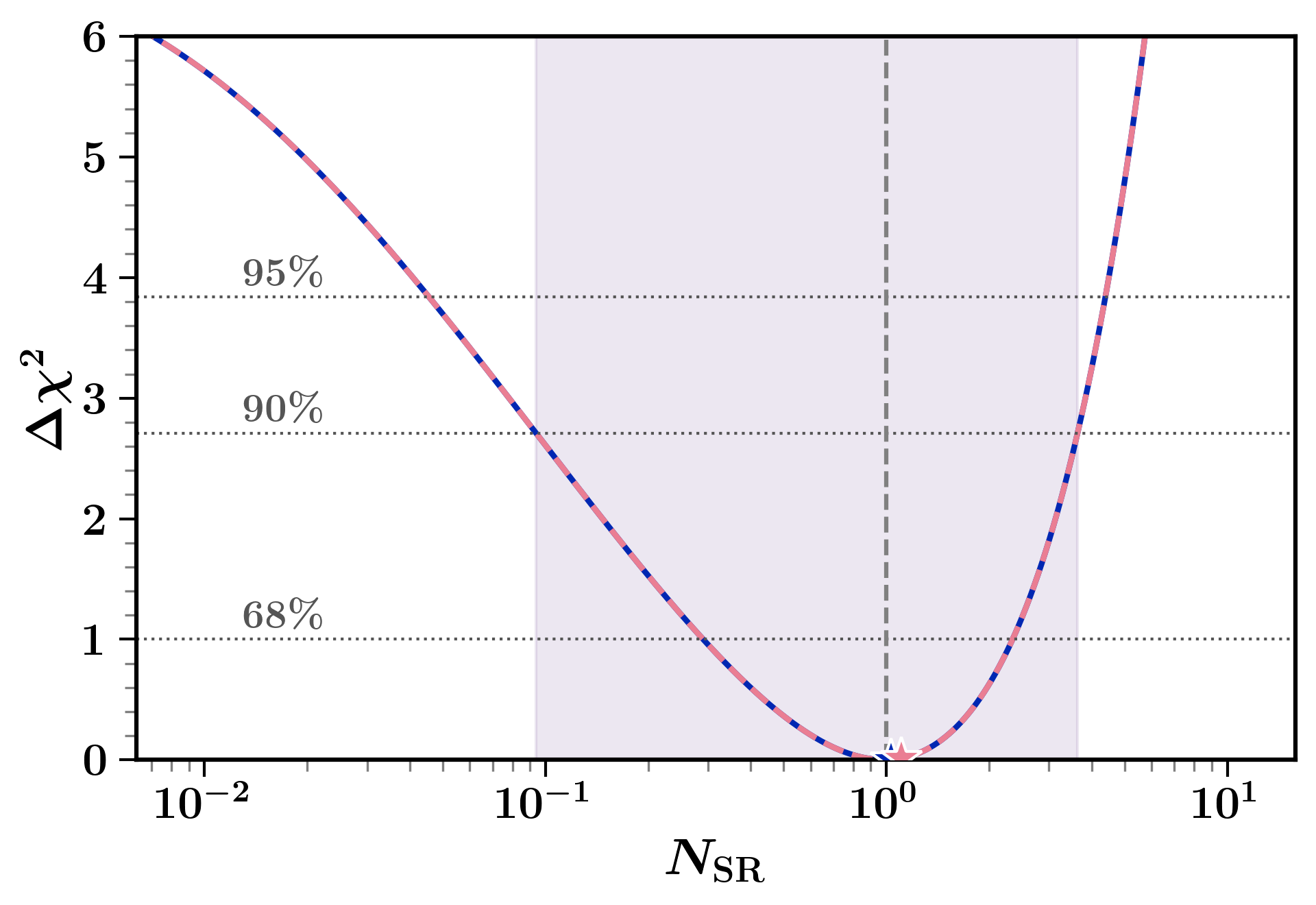}

    \vspace{-0.2cm}
    \makebox[0.48\textwidth][c]{(a)}
    \hfill
    \makebox[0.48\textwidth][c]{(b)}

    \caption{\footnotesize The $\Delta\chi^2$ profile as a function of the expected signal yield $N_{\mathrm{SR}}$ in the high-recoil signal region is shown for the benchmark points considered in Fig.~\ref{fig:3}. Panels (a) and (b) correspond to the model points with DM mass $m_\chi \sim 500$ GeV and $m_\chi = 1$ TeV, respectively. The shaded band indicates the $90\%$ CL interval, $0.1 \lesssim N_{\mathrm{SR}} \lesssim 3.6$, for the expected signal yield.}
    \label{fig:6}
\end{figure*}
Fig.~\ref{fig:6} shows the corresponding profiled $\Delta\chi^2$ as a function of the expected number of signal events, $N_{\mathrm{SR}}$, for the BPs considered in Fig.~\ref{fig:3}. The three horizontal lines indicate the confidence levels (CL) $68\%$, $90\%$, and $95\%$, while the shaded region denotes the CL interval $90\%$, $0.1 \lesssim N_{\mathrm{SR}} \lesssim 3.6$. In both panels, the likelihood is minimized close to $N_{\mathrm{SR}} \simeq 1$, as expected for a single observed event with a negligible background contribution. The nearly identical $\Delta\chi^2$ profiles for the different benchmark points demonstrate that once the recoil spectrum is sufficiently compatible with the observed event, the likelihood constraint is predominantly determined by the overall signal normalization rather than by the detailed spectral shape.

%

\end{document}